\documentclass[prl,twocolumn,showpacs,superscriptaddress,amsmath,amssymb]{revtex4}

\usepackage{graphicx}
\usepackage{dcolumn}
\usepackage{bm}

\begin{document}


\title{Frustration-induced $\eta$ inversion  in the S=1/2 bond-alternating spin chain}

\author{Nobuya Maeshima}
\affiliation{Institute for Molecular Science, Okazaki 444-8585, Japan}
\affiliation{Department of Chemistry, Tokyo Metropolitan University, Hachioji, Tokyo 192-0397, Japan}

\author{Kouichi Okunishi}
\affiliation{Department of Physics, Faculty of Science, Niigata University, Igarashi 2, Niigata 950-2181, Japan}

\author{Kiyomi Okamoto}
\affiliation{Department of Physics, Tokyo Institute of Technology, Meguro-ku, Tokyo 152-8551, Japan}

\author{T\^oru Sakai}
\affiliation{Department of Physics, Tohoku University, Aramaki, Aoba-ku, Sendai 980-8578, Japan}


\begin{abstract}
We study the frustration-induced enhancement of the incommensurate correlation for a bond-alternating quantum spin chain in a magnetic field, which is associated with a quasi-one-dimensional organic compound F$_5$PNN.
We investigate the temperature dependence of the staggered susceptibilities by using the density matrix renormalization group, and then find that the incommensurate correlation becomes dominant in a certain range of the magnetic field.
We also discuss the mechanism of this enhancement on the basis of the mapping to the effective S=1/2 XXZ chain and a possibility of the field-induced incommensurate long range order.
\end{abstract}

\maketitle

The one-dimensional (1D) $S=1/2$ antiferromagnetic bond-alternating spin chain has been an important issue in the condensed matter physics, since it exhibits some typical quantum effects.
The magnetic susceptibility clearly reflects the existence of the dimer spin gap.
A more important aspect is that the Tomonaga-Luttinger (TL) liquid, which is an undoubtedly essential concept in 1D quantum critical systems, is realized  between the vanishing field of the spin gap $H_{c1}$ and the saturation field $H_{c2}$, where the low-energy  behavior of the system is characterized by the TL exponents $\eta_x$ ($\eta_z$) associated with the correlation function for the transverse (longitudinal) staggered mode (see Eq.~(\ref{correlation}))~\cite{sakai-takahashi2,sakai1}.

A good example of such a bond-alternating spin chain is a quasi-1D organic compound F$_5$PNN~\cite{hosokoshi}.
However, a recent precise analysis of  F$_5$PNN  suggests that the frustration effect, which is recently attracting considerable attention, induces various anomalous properties~\cite{goto,izumi1,izumi}. 
In the magnetization process, the bond alternation ratio $\alpha$ (defined in Eq.~(\ref{hamil}))
shows a crossover from $0.4$ in the low field region to $0.5$ in the high field region~\cite{goto}.
Moreover, the temperature dependence of the NMR relaxation rate in a magnetic field exhibits anomalous enhancement of the TL exponent $\eta_z$; 
in contrast to the usual behavior of the TL exponents $\eta_x < \eta_z$ for the simple bond-alternating system~\cite{sakai1},   $\eta_x > \eta_z$ may be realized in a certain range of the magnetic field~\cite{izumi1,izumi}, which we shall call ``$\eta$-inversion'' in this paper.

Motivated by such interesting experimental suggestions of the frustration effect, we study the bond-alternating spin chain with the frustration in a magnetic field $H$: 
\begin{eqnarray}
{\cal H} &=& J\sum_{i}[ \vec{S}_{2i} \cdot \vec{S}_{2i+1} + \alpha   \vec{S}_{2i+1} \cdot \vec{S}_{2i+2} ] \nonumber \\
 &+& J' \sum_i \vec{S}_{i} \cdot \vec{S}_{i+2}  - H\sum_i S^z_i, \label{hamil}
\end{eqnarray}
where $\vec{S}$ is the $S=1/2$ spin operator and $J'$ is the frustrating coupling.
In the following we use the  normalization $J=1$ for simplicity.
Particular importance of this model is that it enable us to capture the interesting physics cooperatively generated by the magnetic field and frustration.
In fact the frustration-induced plateau formation has been studied intensively~\cite{Tone2,totsuka}, and recently a remarkable enhancement of the incommensurate correlation $\eta_x > \eta_z$ has been suggested by the numerical diagonalization analysis in some intermediate magnetic field~\cite{suga}.
However, a systematic investigation of the frustration effect for the TL liquid behavior in the magnetization curve is essentially difficult and a detailed study is clearly desired for  more thorough understanding of its impact on the TL exponents.
In addition, a precise analysis for the $\eta$-inversion  provides an essential view point for the material physics; 
as an interesting consequence of such enhancement of the incommensurate correlation, a novel type of incommensurate order can be induced in the magnetic field through the inter-chain coupling.

In this paper, we reveal the effect of the frustrating coupling $J'$ on the observable quantities, using the finite temperature density matrix renormalization group (DMRG)~\cite{moukouri,wang,shibata}.
We first calculate the magnetization curve and  further investigate the temperature dependence of the staggered susceptibility $\chi_\perp$ ($\chi_\parallel$)  perpendicular (parallel) to the uniform magnetic field.
We then find that the $\eta$-inversion actually occurs in a certain range of the magnetic field, where the gap formation at
 the $M=1/4$ plateau plays a crucial role. Here $M$ is the magnetization per one spin.
Finally we discuss the possibility of the field induced incommensurate order assisted by the frustration. On the basis of the obtained phase diagram, we make a comment on the NMR experiment of F$_5$PNN.

\begin{figure}[h]
\includegraphics[width=6.0cm,clip]{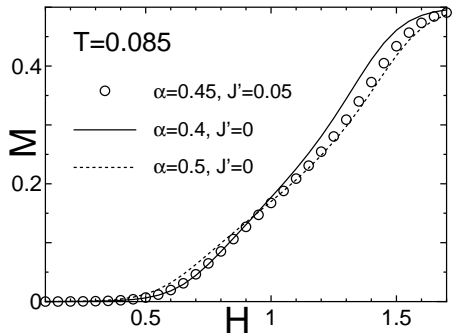}
\caption{Magnetization curves obtained with the DMRG.}
\label{fig:mhcurve}
\end{figure}

In order to analyze the crossover of $\alpha$ in F$_5$PNN, we first calculate the magnetization process by the finite temperature DMRG with the retained bases number $64$.
 Figure~\ref{fig:mhcurve} shows the comparison of the obtained curves at $T=0.085$.
The parameters in the figure correspond to the F$_5$PNN experiment in Ref.~\cite{goto}. 
We can clearly see that the curve of $\alpha=0.45$ with the frustrating coupling $J'=0.05$ explains the crossover of the magnetization process from $\alpha\simeq 0.4$ in the low field region to $\alpha \simeq 0.5$ in the high field region.
This gives a clear evidence of the frustration effect in F$_5$PNN.

Let us next discuss the TL exponents  $\eta_x$ and $\eta_z$, the precise definitions of which are given by power-law decay of the spin-spin correlation functions at zero temperature:
\begin{eqnarray}
\langle S_0^x S_r^x \rangle &\sim& (-1)^r r^{-\eta_x}, \nonumber \\
\langle S_0^z S_r^z \rangle - M^2 &\sim&  \cos(2k_Fr)r^{-\eta_z},
\label{correlation}
\end{eqnarray}
where $k_F=\pi(1/2 - M) $.
It is well known that these exponents provide the essential information for the observable quantities.
For instance, the low temperature behaviors of the staggered susceptibilities are characterized by these exponents~\cite{chitra}:
\begin{equation}
\chi_\perp(T) \sim T^{-(2-\eta_x)} \quad {\rm and} \quad \chi_\parallel(T) \sim T^{-(2-\eta_z)}.
\end{equation}
Since the relation $\eta_x \eta_z =1$ is satisfied for the TL liquid,
the smaller exponent yields the dominant spin fluctuation in low temperatures.
Since the bond-alternating chain usually has $\eta_x < \eta_z$,  $\chi_\perp$ shows the stronger divergence in the $T\to 0$ limit. 
However, if the incommensurate spin correlation along the magnetic field is enhanced, $\chi_\parallel$ can be dominant in the low temperature region.
In order to see how the frustrating coupling $J'$ affects the incommensurate correlation, we directly calculate the staggered susceptibilities $\chi_\perp(T)$ and $\chi_\parallel(T)$ with the finite temperature DMRG~\cite{moukouri,wang,shibata}, which is the most reliable method to investigate an infinitely long chain with the frustration~\cite{maisinger,klumper,maeshima}.
In actual computations, we selectively employ the infinite size and finite size algorithms of the DMRG.
For a calculation of $\chi_\perp$, we perform the infinite size DMRG  with a weak commensurate staggered field along the $x$ direction;  $\chi_\perp$ is obtained as a numerical derivation of the staggered field.
In contrast, it is generally difficult to treat directly the $2k_F$-oscillating magnetic field with the finite temperature DMRG. 
Thus we start with the linear-response formula:
\begin{equation}
\chi_\parallel(T)= \sum_{|r|\le r_{\rm c}} e^{i2k_Fr} \int_{0}^{\beta}d\tau \langle {\cal S}^z(\tau,r) {\cal S}^z(0,0) \rangle,  \label{eq:chidmrg}
\end{equation}
where $\beta$ is the inverse temperature, $r_{\rm c}$ is a cut-off for the real space direction,  and ${\cal S}^z(\tau,r)$ is the spin operator in the Heisenberg representation at an imaginary time $\tau$ and position $r$.
By using the finite size algorithm of the DMRG for the quantum transfer matrix  ${\cal T}$, we calculate the correlation function $\langle {\cal S}^z(\tau,r) {\cal S}^z(0,0) \rangle$.
After obtaining the maximum eigenvalue of $\cal T$ and the corresponding eigenvector $ |\psi_{\rm max}\rangle$,
 we calculate $\langle {\cal S}^z(\tau,r) {\cal S}^z(0,0) \rangle$  as
\begin{equation}
\langle {\cal S}^z(\tau,r) {\cal S}^z(0,0) \rangle = \frac{\langle \psi_{\rm max}| S^z_{\tau,r} {\cal T}^r S^z_{0,0} |\psi_{\rm max}\rangle }{\langle \psi_{\rm max}|{\cal T}^r |\psi_{\rm max}\rangle }. \label{cordmrg}
\end{equation}
Here $S^\alpha_{\tau, r}$ is the spin operator at a position $(\tau,r)$ on the checkerboard lattice obtained via the Suzuki-Trotter decomposition~\cite{ST}.
In the numerical computation, the integral for $\tau$ in Eq.~(\ref{eq:chidmrg}) is replaced by the summation with a finite imaginary-time-slice $\epsilon=\beta/N$, where $N$ is the Trotter number.
In the following results, we have set $r_{\rm c}=200$ and $N=80$, and confirmed the sufficient convergence with respect to $r_{\rm c}$ and $N$.
Here, it should be noted that, especially at $M=1/4$, the consistency of the finite size algorithm can be directly checked with the numerical derivative of the infinite-size DMRG. This is because the renormalization process for the quantum transfer matrix is compatible to the periodicity of $k_F=\pi/2$ at $M=1/4$. We have also  confirmed that the both results are in good agreement with each other.

\begin{figure}[hbt]
\includegraphics[width=6.0cm,clip]{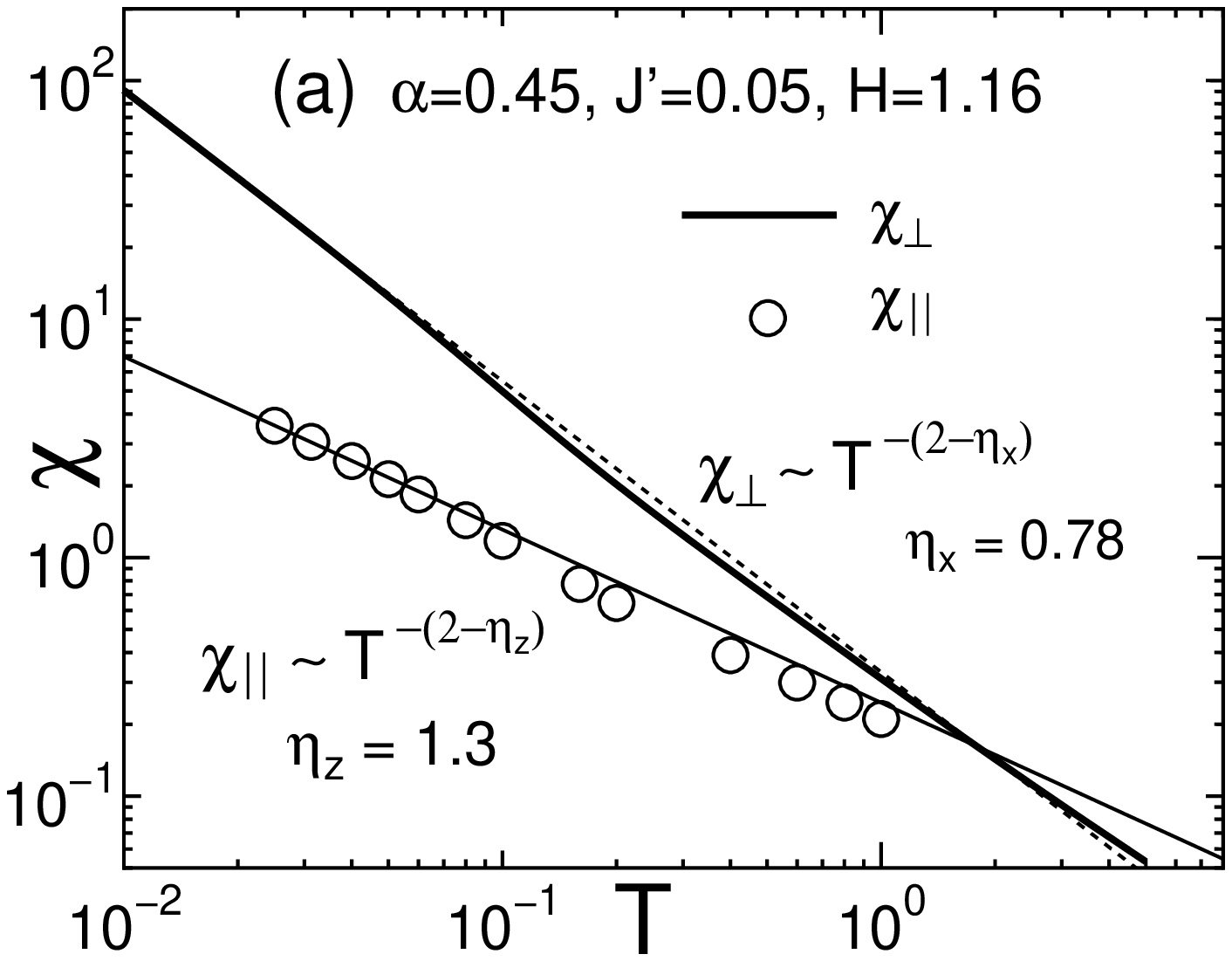}
\includegraphics[width=6.0cm,clip]{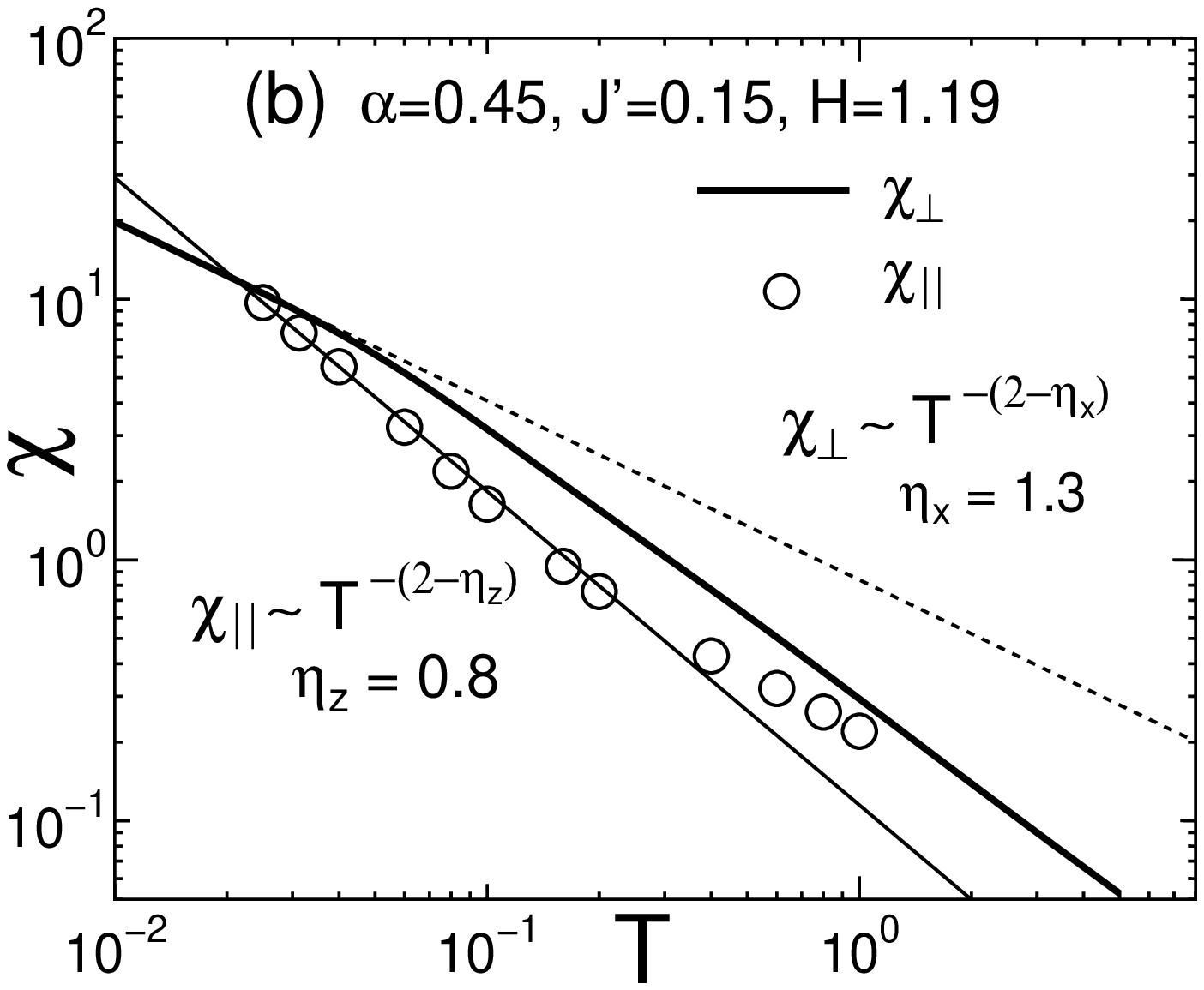}
\caption{Staggered susceptibilities $\chi_\parallel$ and $\chi_\perp$ (a) for $(\alpha,J',H)=(0.45,0.05,1.16)$ and (b) for $(\alpha, J', H)=(0.45,0.15,1.19)$. The dotted (thin solid) lines show the results of power-law fitting for 
$\chi_{\perp} (\chi_\parallel)$. }
\label{fig:kais}
\end{figure}

In Fig.~\ref{fig:kais}, we show comparisons of $\chi_\perp$ and $\chi_\parallel$ for $(\alpha,J',H)=(0.45,0.05,1.16)$ and ($\alpha, J', H)=(0.45,0.15,1.19)$.
In both cases, the corresponding magnetizations at the zero temperature are slightly lower than $M=1/4$.
We can see that $\chi_\perp$ for $J'=0.05$ is always larger than $\chi_\parallel$, implying that the commensurate fluctuation is still dominant.
The estimated exponents are $\eta_x=0.78$ and $\eta_z=1.3$, which yields $\eta_x\eta_z=1.0$.
In contrast, for $J'=0.15$, the susceptibilities clearly show the crossover around $T\sim0.1$;
 as the temperature is decreased below  $T\simeq 0.1$,  the divergence of $\chi_\perp$ is reduced, while that of $\chi_\parallel$ is enhanced.
The TL exponents in the low temperature region are extracted as $\eta_x=1.3$ and $\eta_z=0.8$, which also gives $\eta_x \eta_z \sim 1.0$.
 We can thus verify that the $\eta$-inversion is actually realized by the frustration effect.
In addition, we have found that, as $T \rightarrow 0$, $\chi_\parallel$ in the $M=1/4$ plateau diverges exponentially while $\chi_\perp$ converges to a finite value. 
This fact suggests that the appearance of the plateau gap plays an important role for the $\eta$ inversion.

In order to discuss the microscopic origin of the $\eta$ inversion, let us recall that the Hamiltonian~(\ref{hamil}) around the plateau can be mapped to the effective S=1/2 XXZ chain~\cite{totsuka}, where the anisotropy and magnetic field of the effective model are given by  $\Delta=\frac{1}{2}\frac{2J'+ \alpha}{|2J'-\alpha|}$ and $H_{\rm eff}=H-1-(2J'+\alpha)/4$ respectively.
The exact critical exponents of the XXZ chain in a magnetic field can be derived from the Bethe ansatz integral equation for the dressed charge~\cite{BA}.
Since the $\eta_z$ of the S=1/2 XXZ chain is a monotonous increasing function of $|H_{\rm eff}|$, it is sufficient to investigate $\eta_z$ at $H_{\rm eff}=0$ for the purpose of understanding the appearance of the $\eta$-inversion.
For the XXZ model, the following fact is  well known; as far as $\Delta \le 1$, the XXZ chain is in the critical TL phase with $\eta_z>1$, while, for $\Delta >1$, the excitation gap is opened and, at the same time, $\eta_z <1 $ appears only near the zero magnetic field.  
We therefore find that  {\it  the criterion for $\eta_x > \eta_z$ is equivalent to the one for the gap formation}.
Turning to the original model, we can see that the condition for $\eta_x > \eta_z$ is deduced as  $\alpha /6 <J' < 3\alpha/2$, which is satisfied by the parameters used in Fig.~\ref{fig:kais} (b).
Since the effective anisotropy $\Delta$ becomes large as $J'$ approaches $\alpha /2$, we can also understand that the region $\eta_x >\eta_z$ extends as $J'\to \alpha/2 $.
Although the mapping to the XXZ chain is based on the perturbation theory from the $\alpha,J'\to 0$ limit, the precise study by the level spectroscopy method~\cite{LSM} gives a good estimation for the $1/4$ plateau, where the effective XXZ model picture is basically maintained~\cite{Tone2}.
In fact,  $\eta_z$ of the effective XXZ model for $(\alpha,J',H)=(0.45,0.15,1.19)$ is calculated as $\eta_z=0.735$, which is consistent with the fitting result in Fig.~\ref{fig:kais} (b).
Here we should note that, for $\alpha=0.45$, the $1/4$ plateau at $T=0$ emerges for $0.08<J'<0.5$~\cite{Tone2}, implying that F$_5$PNN is located at a subtle position near the plateau phase boundary.

On the basis of the results mentioned above, let us discuss the effect of the inter-chain interaction $J_{\rm int}$, which induces the 3D long-range order corresponding to the dominant spin correlation~\cite{sakai2}.
Indeed the sharp peak of the specific heat was observed for F$_5$PNN in a magnetic field~\cite{yoshida}.
For the weakly frustrating bond-alternating chain, the antiferromagnetic long-range order is realized for $H_{c1}<H<H_{c2}$.
However, when the strong frustrating coupling $J'$ induces the $\eta$-inversion, the 3D order should also change from the ordinary N\'eel order perpendicular to $H$ into the incommensurate order along $H$ at some field. 
In order to clarify whether such a change of the order occurs or not, we calculate the theoretical phase diagram of the frustrated bond-alternating chain within the inter-chain mean field approximation combined with the DMRG~\cite{klumper,nishiyama};
using $\chi_\parallel(T)$ and $\chi_\perp(T)$ obtained by the finite temperature DMRG,  we can extract the phase boundary from the equation $\chi_\gamma (T_c)=(zJ_{\rm int})^{-1}$~\cite{scalapino}, 
where $z$ is a coordination number and $\gamma \in \perp$ or $\parallel$.
Here we assume that the inter-ladder coupling is weak and is not frustrating.

\begin{figure}[hb]
\includegraphics[width=6.4cm,clip]{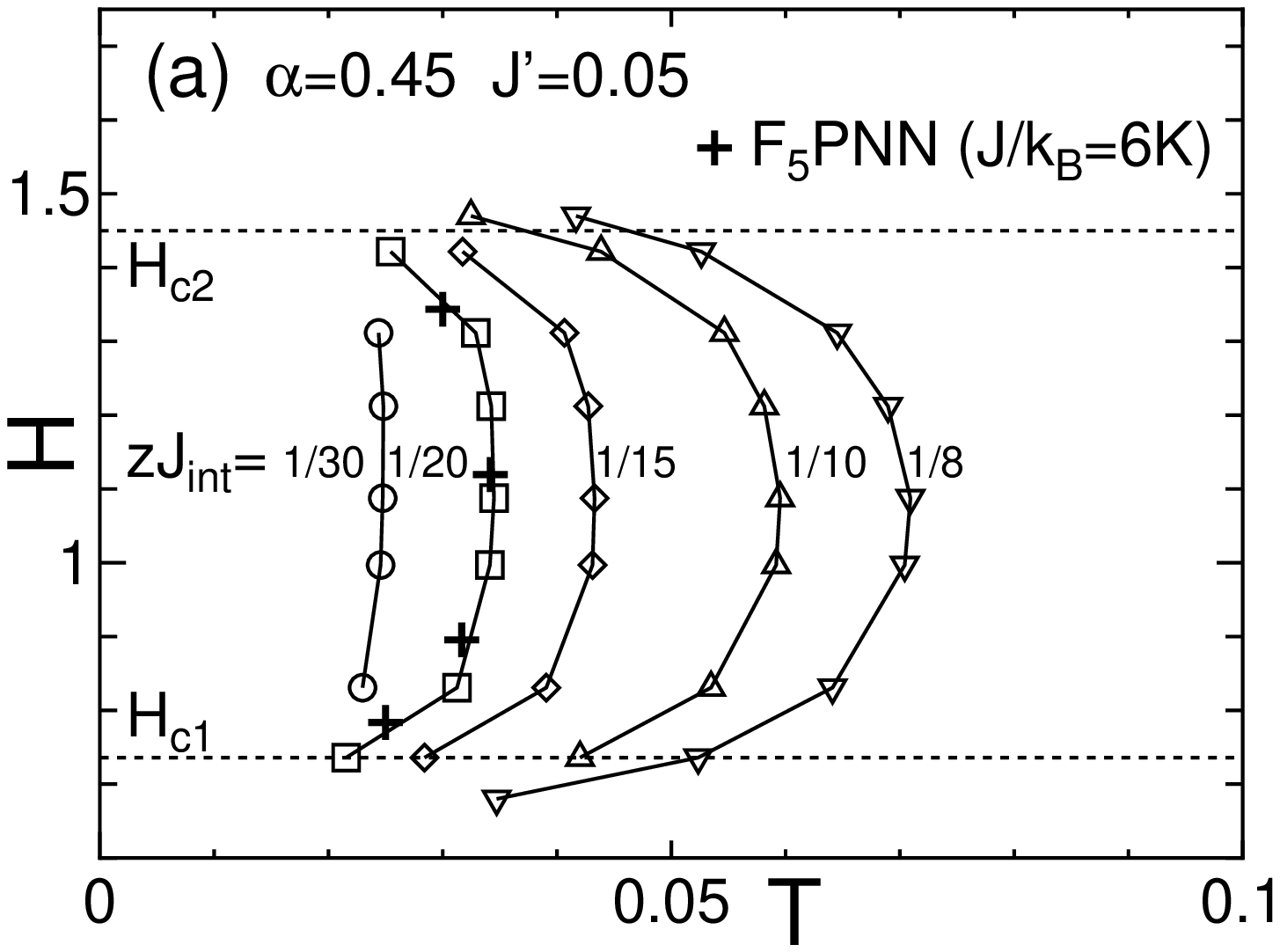}
\includegraphics[width=6.4cm,clip]{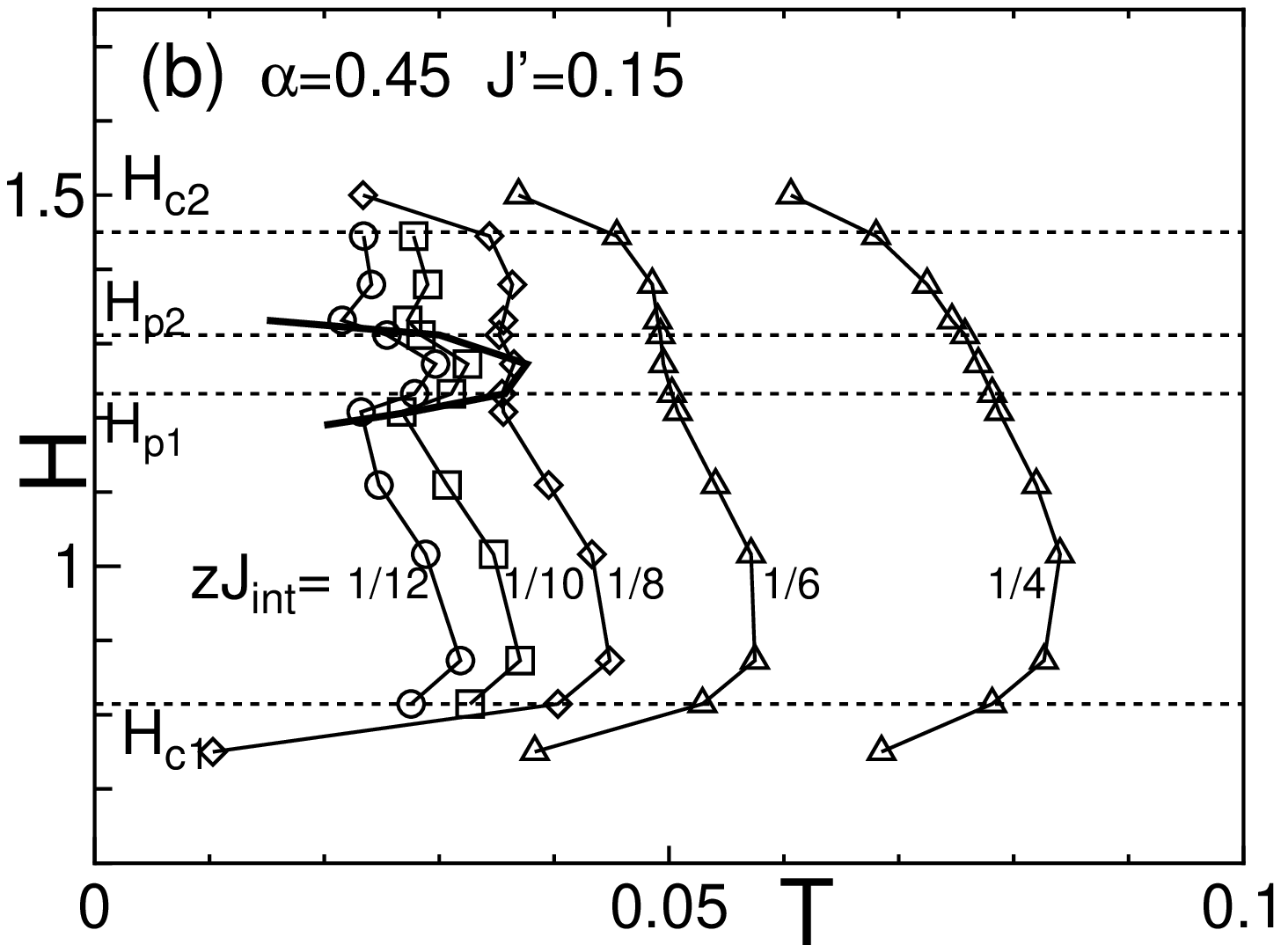}
\caption{ Phase diagram determined by the inter-chain mean field approximation (a) for $(\alpha,J')=(0.45,0.05)$ and (b) for  $(\alpha,J')=(0.45,0.15)$.
Pluses show the critical temperature $T_{\rm c}(H)$ of F$_5$PNN with  $J/k_{\rm B}=6K$.
The bold line in (b) indicates the  $\chi_\parallel=\chi_\perp$ line.
Dotted lines represent the critical fields ($H_{\rm c1}$ and $H_{\rm c2}$), and the lower (upper) edge of the plateau $H_{\rm p1}$ ($H_{\rm p2}$) for the single chain.}
\label{fig:phase}
\end{figure}

In Fig.~\ref{fig:phase}, we show the $H$-$T$ phase diagrams for $\alpha=0.45$ with $J'=0.05$ corresponding to F$_5$PNN, and with $J'=0.15$.
In this figure, the phase boundary for a fixed $zJ_{\rm int}$ is indicated by a line. 
For $\alpha=0.45$ and $J'=0.05$, the transition to the commensurate order always appears, reflecting the fact $\chi_\perp > \chi_\parallel$ in Fig.2 (a).
Moreover, we can see that the phase boundary for F$_5$PNN~\cite{yoshida} is well reproduced by the theoretical curve for $zJ_{\rm int}=1/20$. 
The estimated coupling $J/k_{\rm B}=6$K is also consistent with the one obtained in Ref.~\cite{goto}.
This fact provides another verification for $J'=0.05$ determined from the magnetization curve. 
If we assume $z=4$, the inter-chain coupling is estimated as $J_{\rm int}/k_{\rm B}\simeq0.07$K.
As $J'$ is increased,  the usual commensurate fluctuation is suppressed  around $H\simeq 1$ and, at the same time, the fluctuation associated with $\eta_z$ is increased.
Accordingly  the phase boundary of the commensurate N\'{e}el order is shifted to the low temperature side and, as far as $\chi_{\parallel}= 1/zJ_{\rm int}$ is satisfied, the incommensurate order can appear in the left side of the bold line in Fig. 3 (b), which becomes identical to the commensurate-incommensurate transition line for sufficiently large $J_{\rm int}$.
As was seen before,  the $\eta$ inversion simultaneously appears with the plateau formation.
Thus we can understand that the incommensurate order develops  in the intermediate field region.  

Finally, we want to make a comment on the NMR relaxation rate of F$_5$PNN, for which the theoretical $H$-$T$ phase diagram does not show the $\eta$-inversion. 
Since the temperature region used for the estimation of $\eta_z$ in Ref.~\cite{izumi1,izumi} is close to the transition temperature, the experimentally observed crossover of $\eta_z$ for F$_5$PNN is possibly due to the anomaly originating from the transition to the commensurate N\'{e}el order.
However the $\eta$ inversion certainly occurs in the large $J'$ region.

To summarize, we have studied the bond-alternating spin chain with the frustrating interaction in a magnetic field.
What we want to emphasize in the present study is that  the combination of the magnetic field and frustration gives rise to  various exotic phenomena.
We have actually  shown that the frustration explains the crossover of $\alpha$ in the magnetization curve of F$_5$PNN.
We have also clarified that the $\eta$-inversion of the TL exponents is induced in a certain range of the magnetic field by the frustration.
The mechanism of the $\eta$-inversion is  explained on the basis of the effective XXZ model associated with the 1/4 plateau formation.
Moreover, we discussed the possibility of the field induced incommensurate order.
Although the nature of this transition between commensurate and incommensurate orders is not clear within the mean field theory, we believe that we could elucidate the importance of the incommensurate correlation.
Recently, several frustrated spin systems have been studied, where a low-energy effective XXZ model often works successfully~\cite{totsuka,XXZ,MILA}. In particular the similar enhancement of $\eta_x$ is also reported in Ref.~\cite{MILA}.
This suggests that the mechanism for the $\eta$-inversion can be applicable to a class of realistic frustrating spin systems. 
Then we hope that the present theory stimulates further researches on such interesting physics caused by the incommensurate correlation.

We would like to thank S. Suga, T. Suzuki, Y. Yoshida, K. Izumi, Tsuneaki Goto and Takao Goto for fruitful discussions. This work was partially supported by Grants-in-Aid for Scientific Research on Priority Areas (B),  for Scientific Research (C) (No.14540329), and for Creative Scientific Research from the Ministry of Education, Culture, Sports, Science and Technology, Japan.

\end{document}